\documentclass[a4paper]{article}
\usepackage[utf8]{inputenc}
\usepackage{float} 
\usepackage[margin=25mm]{geometry}
\usepackage{amsmath}
\usepackage{amsfonts}
\usepackage{amssymb}
\usepackage{graphicx}
\usepackage{authblk}
\usepackage{comment}
\usepackage[style=chem-angew]{biblatex}
\title{Semiclassical trajectories in the double-slit experiment}
\author{Hector H. Hernandez Hernandez\footnote{hhernandez@uach.mx}   , Carlos R. Javier Valdez\footnote{a310861@uach.mx} }
\affil{Universidad Autonoma de Chihuahua, Facultad de Ingenieria, Nuevo Campus Universitario, Chihuahua 31125, Mexico}
\date{\today}

\begin{document}

\maketitle

\begin{abstract}
    We provide a semiclassical description of the double-slit experiment based on momentous quantum mechanics, where the implementation of canonical variables facilitate the derivation of the equations of motion for the system. We show the evolution of individual particles and their semiclassical trajectories, collectively reproduce the well-known quantum interference pattern. It is found  that the non-crossing rule for trajectories, present in Bohmian mechanics, is not required under our treatment. We are able to follow classical configuration variables from this semiclassical scheme, and discuss substantial differences between our description and the Bohmian perspective.
\end{abstract}

Keywords: Semiclassical quantum mechanics. Double-slit experiment.

PACS: 03.65.Sq, 03.65.w, 03.65.Xp

\hspace{5pt}

\section{Introduction}
One of the most prominent phenomenon in modern physics is the double-slit experiment because it alone shows most of the important features of quantum mechanics, such as the wave-particle duality, wave function collapse, existence of non locality, a way to verify the Born-rule, on different setups \cite{PhysRevA.60.1530,PhysRevA.95.042129,PhysRevLett.75.1252,PhysRevA.95.042129,PhysRevA.46.R17,PhysRevLett.95.040401}. These features can be explained within the quantum mechanics framework. However, there are some classical properties, such as the time of flight of the particles that cross through the double slit, that standard quantum mechanics cannot explain or compute by the corresponding expectation value \cite{PhysRevA.77.014102}. Standard quantum mechanics lacks a particle-like description, in the classical sense, and so there exist alternative proposals or interpretations \cite{PhysRev.85.166,PhysRevA.87.014102}, being Bohmian mechanics the most complete \cite{PhysRevLett.110.060406}. In spite of the great concordance that Bohmian mechanics has with standard quantum mechanics, it presents important differences,  such as the existence of particle trajectories evolving and interacting with a quantum potential \cite{holland_1993}, or the interpretation of the superposition principle \cite{Nassar2017,GHOSE2001205,GROSSING2012421}. These differences, however, have implications in the interpretation of the results on the double-slit experiment, such as the reality of trajectories \cite{englert1992surrealistic}.\\
In this paper we study the double-slit experiment under a semiclassical approach, where we obtain the well-known interference pattern, as well as particle trajectories. These semiclassical trajectories are constrained due to the uncertainty relation of variables on an extended phase space, being in more concordance with the usual interpretation of quantum mechanics. This semiclassical approach is based on an extension of Ehrenfest's theorem \cite{doi:10.1142/S0129055X06002772,PhysRevA.99.042114,PhysRevA.98.063417,doi:10.1063/1.4748550} and has been successfully employed to study systems such as quantum tunneling, quantum pendulum and quantum cosmology \cite{doi:10.1063/1.4748550,10.1142,PhysRevA.98.063417,PhysRevD.84.043514}.

\section{Effective dynamics of quantum systems: general definition}
The dynamics of expectation values of observables in quantum systems can be described by employing the  momentous formulation of quantum mechanics. In this scheme, the evolution of the system is obtained by means of a set of effective equations for expectation values of observables and their associated momenta, allowing us to understand how the classical dynamics of the system is corrected by including quantum effects.

The effective equations of motion are obtained by evaluating Poisson brackets between variables and the quantum corrected Hamiltonian, $\dot f= \{ f,H_{Q} \}$, where $H_{Q}$ is the classical Hamiltonian plus quantum corrections. Expectation values of position, momentum and quantum dispersions are considered classical variables.
Expectation values of dispersions and momenta are defined by
\begin{equation}
\label{Effective dynamical variables}
    G^{a,b}:=\langle (\hat x -x)^{a}(\hat p -p)^{b}\rangle_{\textrm{Weyl},}\quad a+b\geq 2, 
\end{equation}
whereas the quantum corrected Hamiltonian reads
\begin{equation}
\label{Quantum corrected Hamiltonian}
    \langle \hat H \rangle:=H_{Q}=H(x,p)+\sum_{a,b} \frac{1}{a!b!}\frac{\partial^{a+b}H}{\partial x^{a}\partial p^{b}}G^{a,b} 
\end{equation}
where $x:=\langle \hat x\rangle$, $p:=\langle \hat p\rangle$, and the operators are Weyl ordered. Usual quantum fluctuations can be directly identified: $\Delta x^{2}=G^{2,0}$ and $\Delta p^{2}=G^{0,2}$.

In general, an infinite number of coupled equations is obtained from (\ref{Quantum corrected Hamiltonian}), providing a complete description of the quantum system. Recently, it was obtained a generalization of this effective method in terms of explicitly canonical variables ($s,p_{s},U$) \cite{PhysRevA.98.063417} 
\begin{equation}
\label{Canonical variables s and p_{s}}
    s=\sqrt{G^{2,0}}, \quad\quad p_{s}=\frac{G^{1,1}}{\sqrt{G^{2,0}}}
\end{equation}
\begin{equation}
\label{U}
    U=G^{2,0}G^{0,2}-(G^{1,1})^{2}
\end{equation}
In terms of these variables the Hamiltonian (\ref{Quantum corrected Hamiltonian}) can be rewritten as
\begin{equation}\label{Hamiltonian with effective potential}
\begin{split}
  H_{Q}&=\frac{p_{x}^{2}+p_{s}^{2}}{2m}+V(x)+\frac{U}{2ms^{2}}+\sum_{a} \frac{1}{a!}\frac{\partial^{a}V}{\partial x^{a}}G^{a,0}\\
\end{split} 
\end{equation}
or in a more compact form
\begin{equation}
\label{Canonical Hamiltonian}
\begin{split}
  H_{Q}&=\frac{p_{x}^{2}+p_{s}^{2}}{2m}+\frac{U}{2ms^{2}}+\frac{1}{2}(V(x+s)+V(x-s)).
\end{split} 
\end{equation}
The extension to several degrees of freedom can be obtained in a similar way, with a generalization of (\ref{Effective dynamical variables})
\begin{equation}
\label{Generalized effective dynamical variables}
\begin{split}
    G^{a_{1},b_{1},\dots,a_{k},b_{k}}:=&\langle (\hat x_{1} -x_{1})^{a_{1}}(\hat p_{1} -p_{1})^{b_{1}}\cdots(\hat x_{k} -x_{k})^{a_{k}}(\hat p_{k} -p_{k})^{b_{k}}\rangle_{\textrm{Weyl}} 
\end{split}    
\end{equation}
where $a_{i}+b_{i}\geq 2$. In this case the general quantum-corrected Hamiltonian  is
\begin{equation}
\label{Generalized Quantum corrected Hamiltonian}
\begin{split}
    H_{Q}:=&\sum_{a_{1},b_{1}}^{\infty}\cdots\sum_{a_{k},b_{k}}^{\infty}\frac{1}{a_{1}!b_{1}!\cdots a_{k}!b_{k}!     }\frac{\partial^{ a_{1}+b_{1}+\cdots+a_{k}+b_{k}}H}{\partial x_{1}^{a_{1}}\partial p_{1}^{b_{1}}\cdots\partial x_{k}^{a_{k}}\partial p_{k}^{b_{k}}}G^{a_{1},b_{1},\dots,a_{k},b_{k}}
\end{split}
\end{equation}

\section{Effective quantum Hamiltonian for the double-slit experiment}
The configuration for the double-slit experiment consists of the following: a source of particles, such as electrons, atoms, etc., a double slit placed in front of it, and a screen that records the position of particles that reach it.

\begin{figure}[h]
\centering
\includegraphics[width=8cm]{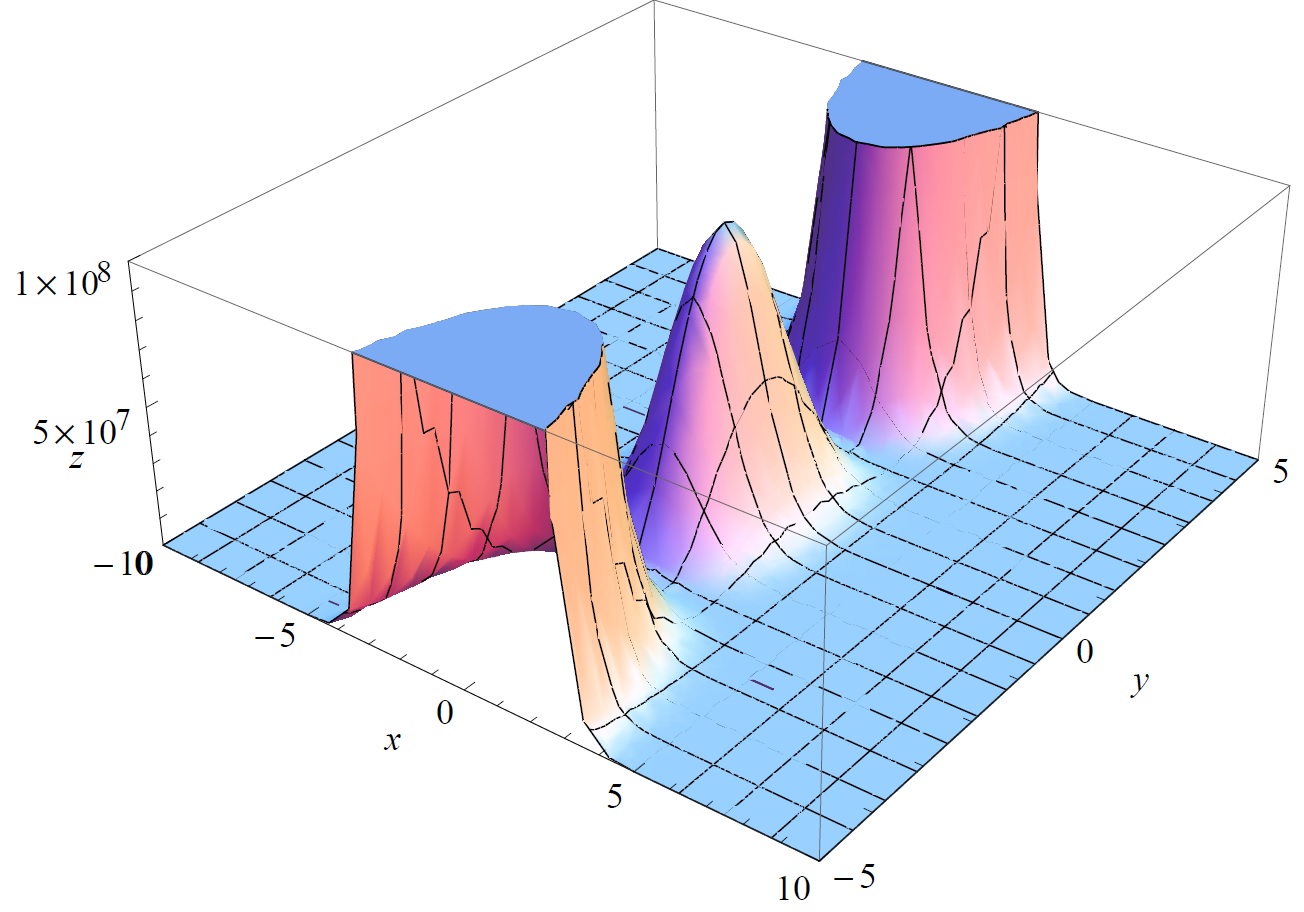}
\caption{Double slit-potential with $m=1$, $\omega=10000$, $V_{o}=10^{7}$ and $\alpha=1.5$.}
\label{fig:Double slit graph}
\end{figure}

We use the double-slit potential function shown in Fig. \ref{fig:Double slit graph}, introduced in \cite{prezhdo2006quantized} 
\begin{equation}
\label{Double slit potential}
    V(x,y)=\left( V_{o}-\frac{1}{2}m\omega^{2}y^{2}+\frac{m^{2}\omega^{4}y^{4}}{16V_{o}}\right)e^{-(x/ \alpha)^{2}}
\end{equation}
The trajectories of the particles are confined to the xy-plane, thus we set $k=2$ in the general quantum corrected Hamiltonian (\ref{Complete equation of the effective Hamiltonian})
\begin{equation}\label{Complete equation of the effective Hamiltonian}
\begin{split}
H_{Q}&=\frac{p_{x}^{2}}{2m}+\frac{p_{y}^{2}}{2m}+\frac{1}{2m}G^{0,2,0,0}+\frac{1}{2m}G^{0,0,0,2}+V(x,y)+\sum_{a_{2}=2}^{\infty}\frac{1}{a_{2}!}\frac{\partial^{ a_{2}}V(x,y)}{\partial y_{2}^{a_{2}}}G^{0,0,a_{2},0}\\
&+\sum_{a_{1}=2}^{\infty}\frac{1}{a_{1}!}\frac{\partial^{ a_{1}}V(x,y)}{\partial x_{1}^{a_{1}}}G^{a_{1},0,0,0}
\end{split}
\end{equation}
In terms of canonical variables the Hamiltonian reads
\begin{equation}
\label{New Hamiltonian for all orders}
\begin{split}
H_{Q}&=\frac{p_{x}^{2}+p_{sx}^{2}}{2m}+\frac{p_{y}^{2}+p_{sy}^{2}}{2m}+\frac{U}{2ms_{x}^{2}}+\frac{U}{2ms_{y}^{2}}+\frac{1}{4}\sum_{i,j=1}^{2} V(x+(-1)^{i}s_{x},y+(-1)^{j}s_{y})
\end{split}
\end{equation}
The corresponding equations of motion are
\begin{equation}
\label{Equations of the double slit}
    \begin{split}
        &\dot x=\frac{p_{x}}{m}, \quad\quad\quad\quad\quad \dot s_{x}=\frac{p_{sx}}{m},\\
        &\dot y=\frac{p_{y}}{m},
        \quad\quad\quad\quad\quad\dot s_{y}=\frac{p_{sy}}{m},\\
        &\dot p_{x}=\frac{1}{2\alpha^{2}}\Big(2V_{o}-m\omega^{2}(y^{2}+s_{y}^{2})+\frac{m^{2}\omega^{4}}{8 V_{o}}(y^{4}+6y^{2}s_{y}^{2}+s_{y}^{4})\Big)\Big((x-s_{x})e^{-(\frac{x-s_{x}}{\alpha})^{2}}+(x+s_{x})e^{-(\frac{x+s_{x}}{\alpha})^{2}}\Big), \\
        &\dot p_{sx}=\frac{U}{ms_{x}^{3}}+\frac{1}{2\alpha^{2}}\Big( 2V_{o}-m\omega^{2}(y^{2}+s_{y}^{2})+\frac{m^{2}\omega^{4}}{8V_{o}}(y^{4}+6y^{2}s_{y}^{2}+s_{y}^{4})\Big)\Big( (x+s_{x})e^{-(\frac{x+s_{x}}{\alpha})^{2}}-(x-s_{x})e^{-(\frac{x-s_{x}}{\alpha})^{2}}\Big), \\
        &\dot p_{y}=\Big(\frac{m\omega^{2}y}{2}-\frac{m^{2}\omega^{4}}{8Vo}(s_{y}^{3}+3ys_{y}^{2})\Big)\Big(e^{-(\frac{x+s_{x}}{\alpha})^{2}}
        +e^{-(\frac{x-s_{x}}{\alpha})^{2}}\Big), \\
        &\dot p_{sy}=\frac{U}{ms_{y}^{3}}+\Big( \frac{m\omega^{2}s_{y}}{2}-\frac{m^{2}\omega^{4}}{8V_{o}}(s_{y}^{3}+3y^{2}s_{y}) \Big)\Big(e^{-(\frac{x+s_{x}}{\alpha})^{2}}+e^{-(\frac{x-s_{x}}{\alpha})^{2}}\Big)
    \end{split}
\end{equation}

\section{Semiclassical description of the double-slit experiment}
\label{section: Semiclassical description}
As we mentioned above, the dynamics described by the system (\ref{Equations of the double slit}), represents semiclassically the behavior of the quantum system under consideration, and thus it is constrained by Heisenberg's uncertainty (\ref{Heisenberg uncertainty principle}). This relation allows us to establish initial conditions for effective variables $(G^{a,b})$. We can impose, for instance, that quantum variables saturate the lower bound in (\ref{Heisenberg uncertainty principle})
\begin{equation}
\label{Heisenberg uncertainty principle}
    G^{2,0}_{o}G^{0,2}_{o}-(G^{1,1}_{o})^{2}=\frac{\hbar^{2}}{4}
\end{equation}
We propose an initial Gaussian state
\begin{equation}
\label{Initial Gaussian state}
    \begin{split}
        \Psi_{o}=\frac{1}{(2\pi\sigma_{o}^{2})^{1/4}}\text{exp}\Big( -\frac{(x-x_{o})^{2}}{4\sigma_{o}^{2}}+i\frac{p_{o}}{\hbar}(x-x_{o}) \Big).
    \end{split}
\end{equation}
from which quantum fluctuations $G_{o}^{2,0}=\Delta x_{o}^{2}$, $G_{o}^{0,2}=\Delta p_{o}^{2}$ and $G_{o}^{1,1}=\Delta x_{o}p_{o}$ can be obtained by computing usual expectation values, for example
\begin{equation}
    \begin{split}
        G_{o}^{1,1}&=\langle \Psi_{o}|(\hat{x}-x_{o})(\hat{p}-p_{o})|\Psi_{o}\rangle.\\
    \end{split}
\end{equation}
In this particular case, we obtain $G_{o}^{1,1}=0$, and (\ref{Heisenberg uncertainty principle}) becomes
\begin{equation}
\label{Inequality with G11=0}
    \begin{split}
       G^{2,0}_{o}G^{0,2}_{o}=\frac{\hbar^{2}}{4}  
    \end{split}
\end{equation}
For the two dimensional case we obtain two sets of initial conditions computed in a similar way, that is, (\ref{Heisenberg uncertainty principle}) is replaced by 
\begin{equation}
\label{Heisenberg equations for 2 dimensions}
\begin{split}
   &G^{2,0}_{io}G^{0,2}_{io}-(G^{1,1}_{io})^{2}=\frac{\hbar^{2}}{4},\quad i\in [x,y],
\end{split}    
\end{equation}
and the initial gaussian state is given by 
\begin{equation}
\label{two dimensional gaussian state}
    \begin{split}
        \Psi_{o}(x,y)=\Psi_{o}(x)\Psi_{o}(y).
    \end{split}
\end{equation}
Once again $G_{io}^{1,1}=0$. 

We are interested in describing the evolution of the system in terms of canonical variables $(s,p_{s},U)$. From (\ref{Canonical variables s and p_{s}})  and (\ref{U}) we obtain

\begin{equation}
\label{G02 equation}
\begin{split}
    G^{0,2}&=\frac{1}{s^{2}}(U+p_{s}^{2}s^{2}).\\
\end{split}
\end{equation}
From $G_{o}^{1,1}=0$ we see that $p_{s_o}=0$ and (\ref{G02 equation}) reduces to
\begin{equation}
\label{G02 initial}
\begin{split}
    G^{0,2}_{o}&=\frac{U}{s_{o}^{2}}.
\end{split}
\end{equation}
In the two dimensional regime, (\ref{G02 initial}) is replaced by 
\begin{equation}
\label{G02io}
    \begin{split}
    G^{0,2}_{io}=\frac{U}{s^{2}_{io}},\quad p_{s_io}=0,\quad  i\in[x,y]. 
    \end{split}
\end{equation}
From these expressions we obtain initial values for canonical variables $s_{i}$. 

\subsection{Semiclassical trajectories}
\label{subsection: Semiclassical trajectories}
Equations (\ref{Equations of the double slit}) constitutes a nonlinear system of coupled differential equations, for which no analytical solution exists.
\begin{figure}[h]
\centering
\includegraphics[width=7.5cm]{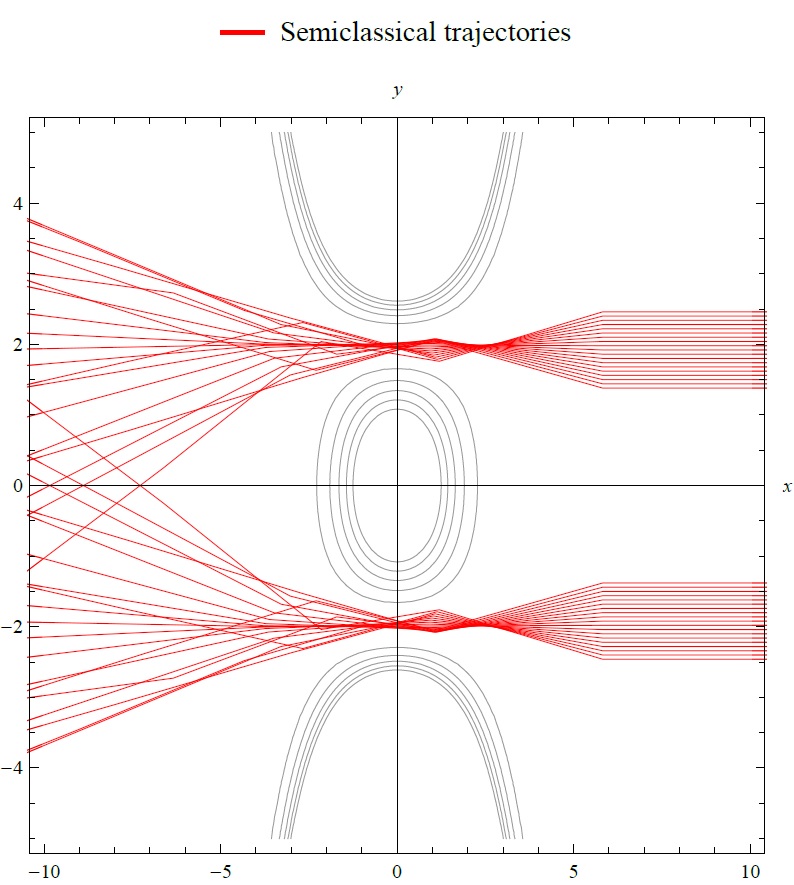}
\caption{Semiclassical trajectories for the double-slit experiment.}
\label{fig: Semiclassical trajectories}
\end{figure}
Initial conditions for classical variables $(x,y,p_{x},p_{y})$ can be fixed from the configuration of the experiment, as in \cite{PhysRevA.87.014102,doi:10.1119/1.16104}. These conditions are modified mainly by different experimental arrangements depending on the type of particles employed \cite{PhysRevA.46.R17,PhysRevA.77.014102,doi:10.1119/1.16104}, the material used as the barrier \cite{PhysRevA.60.1530}, and the properties under study in the experiment \cite{PhysRevLett.95.040401}. This variety of configurations establish different initial conditions, whose values change slightly from those in \cite{PhysRevLett.75.1252,PhysRevA.46.R17,PhysRevLett.95.040401,PhysRevA.77.014102,doi:10.1119/1.16104}.
We employ initial conditions for classical variables as in \cite{PhysRevA.77.014102}, and for canonical variables $(s_{x},p_{sx},s_{y},p_{sy})$, as discussed in section \ref{section: Semiclassical description}, we set
\begin{equation}
\label{Complete initial conditions for the double slit experiment}
\begin{split}
    &x_{o}=400,\quad y_{o}\in [-4,4],\quad p_{xo}=-5000,\quad p_{yo}=0,\\
    &p_{sxo}=0,\quad p_{syo}=0,\quad s_{xo}=0.2,\quad s_{yo}=0.2,\quad U=0.25
\end{split}
\end{equation}
Feeding these data in (\ref{Equations of the double slit}) one can solve numerically. We show below several behaviors for particle evolution.

Semiclassical trajectories for individual particles crossing the potential barrier are shown in Fig. \ref{fig: Semiclassical trajectories}. Notice that some particles, coming from upper half of the barrier, can be detected in the lower half of the plane and vice versa. The recording screen is placed at $x=-350$. 

\begin{figure}[h]
    \centering
    \includegraphics[width=7.5cm]{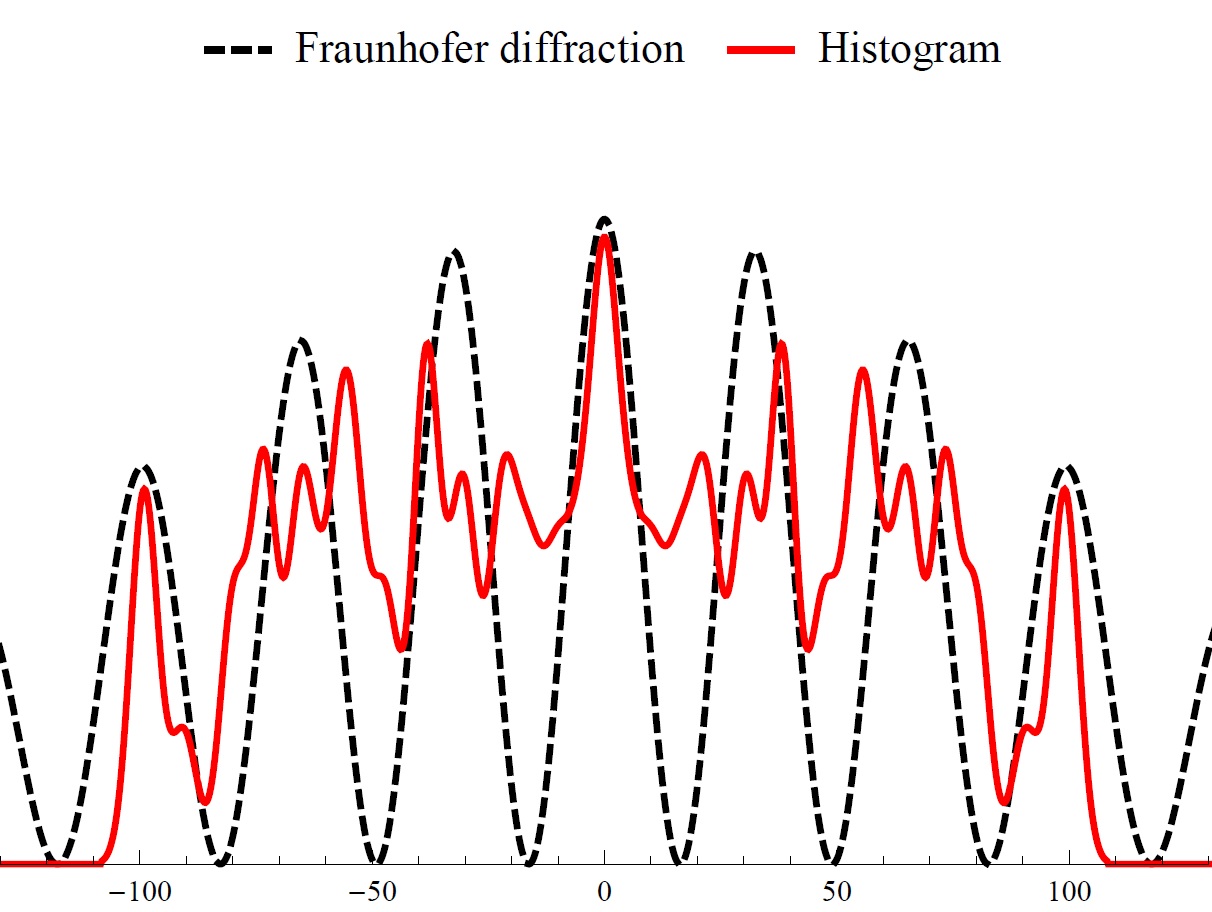}
    \caption{Frequency curve of particles arriving at $x=-350$ (red line), and the Fraunhofer diffraction pattern (black dashed line) which shows clearly that the maximum fringe is located where most of the lines in Fig. \ref{fig: Semiclassical trajectories} are overlapped.}
    \label{fig:Histogram}
\end{figure}

Most particles are detected around the symmetry line, displaying the behavior shown in Fig. \ref{fig:Histogram} 
\begin{figure}[h]
    \centering
    \includegraphics[width=7.5cm]{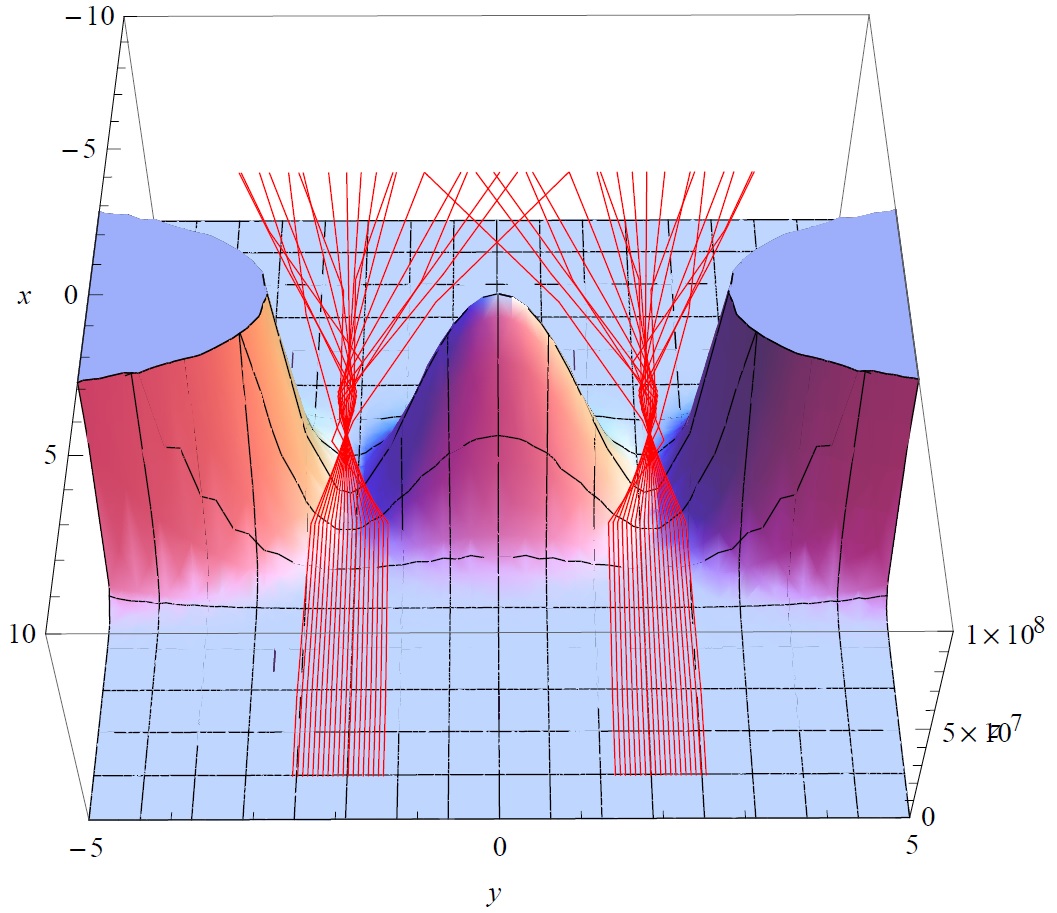}
    \caption{Semiclassical 3D trajectories for the double-slit potential with $E=25\cdot 10^{6}$}
    \label{fig:Semiclassical trajectories 3D}
\end{figure}

\subsection{Trajectories with quantum dispersions and new insights}
Momentous quantum mechanics allows obtaining
of semiclassical trajectories, and it also provides a way to compute the evolution of dispersions  (\ref{Heisenberg uncertainty principle}). Quantum uncertainty must be taken into account, and thus the position of individual particles is computed in the following way
\begin{equation}
\label{Uncertainty belt}
   x(t)\pm s_{x}(t),\quad \text{and}\quad y(t)\pm s_{y}(t).
\end{equation}

\begin{figure}[h]
    \centering
    \includegraphics[width=7.5cm]{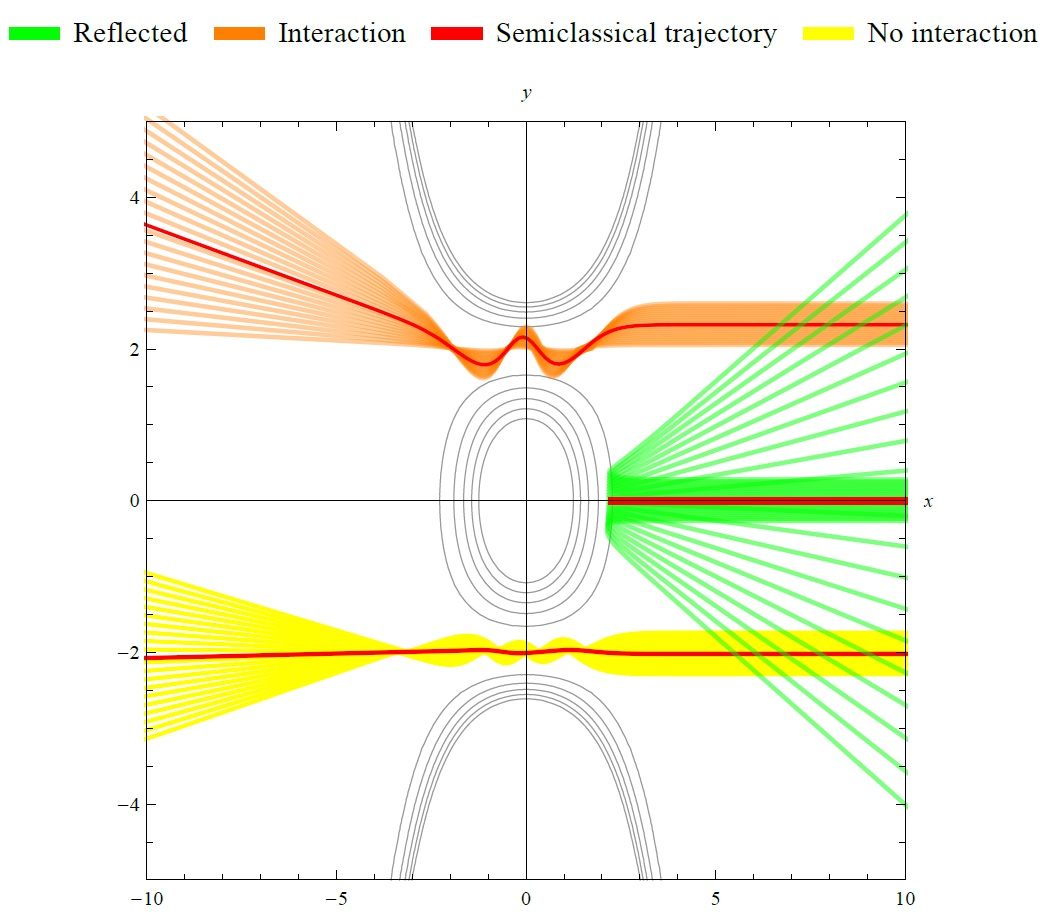}
    \caption{Trajectories in the double-slit experiment: particles being reflected (green area), particles strongly interacting with the potential (orange area), and particles with weak interaction (yellow area)}
    \label{fig: Trajectories behavior}
\end{figure}

\begin{figure}[h]
    \centering
    \includegraphics[width=6.5cm]{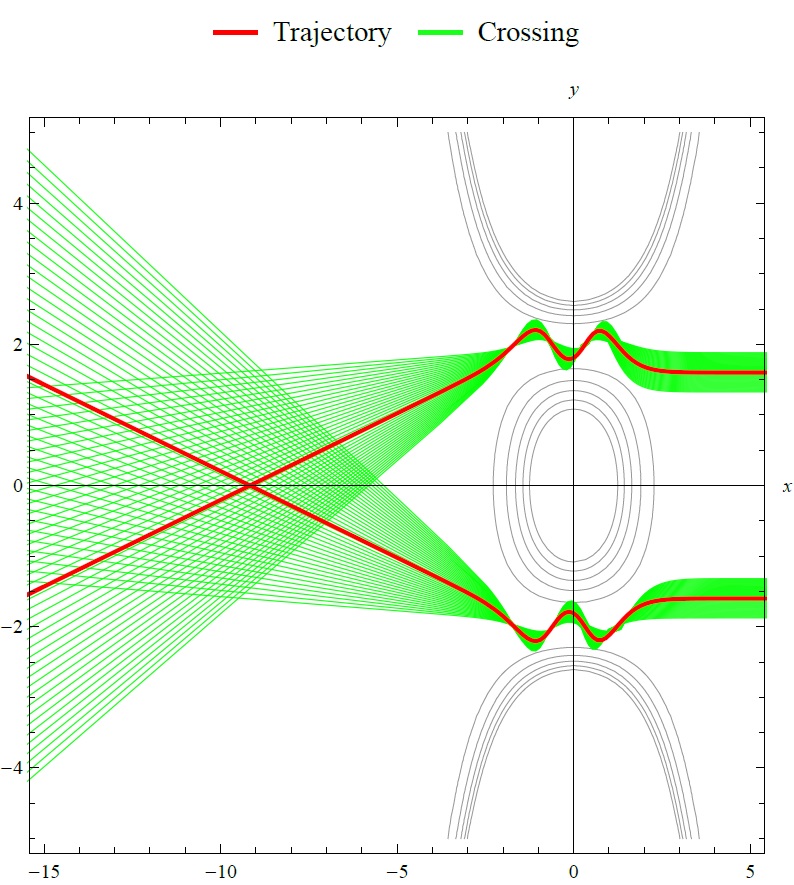}
    \caption{Quantum uncertainty (green area) for trajectories (red line), crossing after the barrier.}
    \label{fig:Semiclassical trajectories together with its uncertainty belt.}
\end{figure}

\begin{figure}[h]
    \centering
    \includegraphics[width=6.5cm]{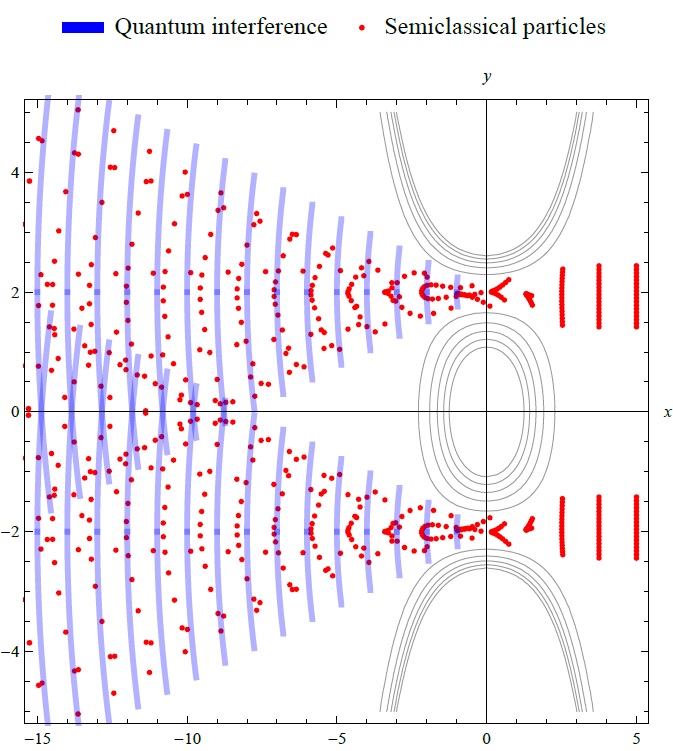}
    \caption{Particles (red dots) clustering in wavefronts (blue line).}
    \label{fig:Comparison between quantum interference}
\end{figure}

Several particle trajectories and their dispersions are plotted in Fig. \ref{fig: Trajectories behavior}, where the uncertainty belt is also shown (\ref{Uncertainty belt}).
Uncertainty for crossing trajectories, as mentioned above, is shown in Fig. \ref{fig:Semiclassical trajectories together with its uncertainty belt.}.
In Fig. \ref{fig:Comparison between quantum interference}, point particles display their wave behavior: one can see this as the overlapping of front waves coming out of the slits. 

\section{Discussion}
There exist different explanations for the double-slit experiment in the literature. According to standard quantum mechanics, particles can cross both slits at the same time, even in a one-by-one setup \cite{doi:10.1119/1.16104}. However, this takes into account only the interference pattern displayed on the screen, it does not consider which slit the particle has crossed. Which-way experiments \cite{PhysRevA.95.042129} attempt to determine that: one wishes to establish which slit the particle has crossed, and in some cases, particles coming from the upper half of the double slit can be recorded in the lower part of the screen, and vice versa \cite{englert1992surrealistic}. 

There is no classical description for this. However, Bohmian mechanics, which offers a way to study quantum systems without abandoning classical concepts, such as trajectories, velocities, positions, may provide a way to explain exactly that. It is capable of reproducing the same statistical predictions as standard quantum mechanics. However, trajectories in this scheme display what is called the "non-crossing rule" \cite{Nassar2017}, which simply states that particles crossing the upper slit do not cross to the lower half of the screen, and vice versa. This behavior is in direct conflict with the results obtained in the experiments. 

These features can be explained directly in our model. Particles arriving at the barrier interact with the potential, cross the slits and then follow semiclassical trajectories. Some of them remain in their half plane and others cross to the opposite side on the screen, as shown in figures \ref{fig: Semiclassical trajectories},\ref{fig: Trajectories behavior},\ref{fig:Semiclassical trajectories together with its uncertainty belt.},\ref{fig:Comparison between quantum interference}. Our results are in no way classical, quantum dispersions are also dynamical and modify effectively the classical behavior of particles and their trajectories, as shown in Fig.\ref{fig:Semiclassical trajectories together with its uncertainty belt.}. This can be interpreted in the following way: at the barrier the dispersion get squeezed by interacting with the potential, this provokes a significant modification of their classical behavior through quantum back reaction, altering the original trajectory and allowing deflection to the opposite half plane in some cases.
By considering the individual position of particles, we observe how their wave nature is reproduced by this effective dynamical evolution. In this picture, interference can be understood as particles clustering in a small region of space, behaving as if they were traveling on wave fronts, as shown in Fig. \ref{fig:Comparison between quantum interference}. 
\begin{figure}[h]
    \centering
    \includegraphics[width=6.5cm]{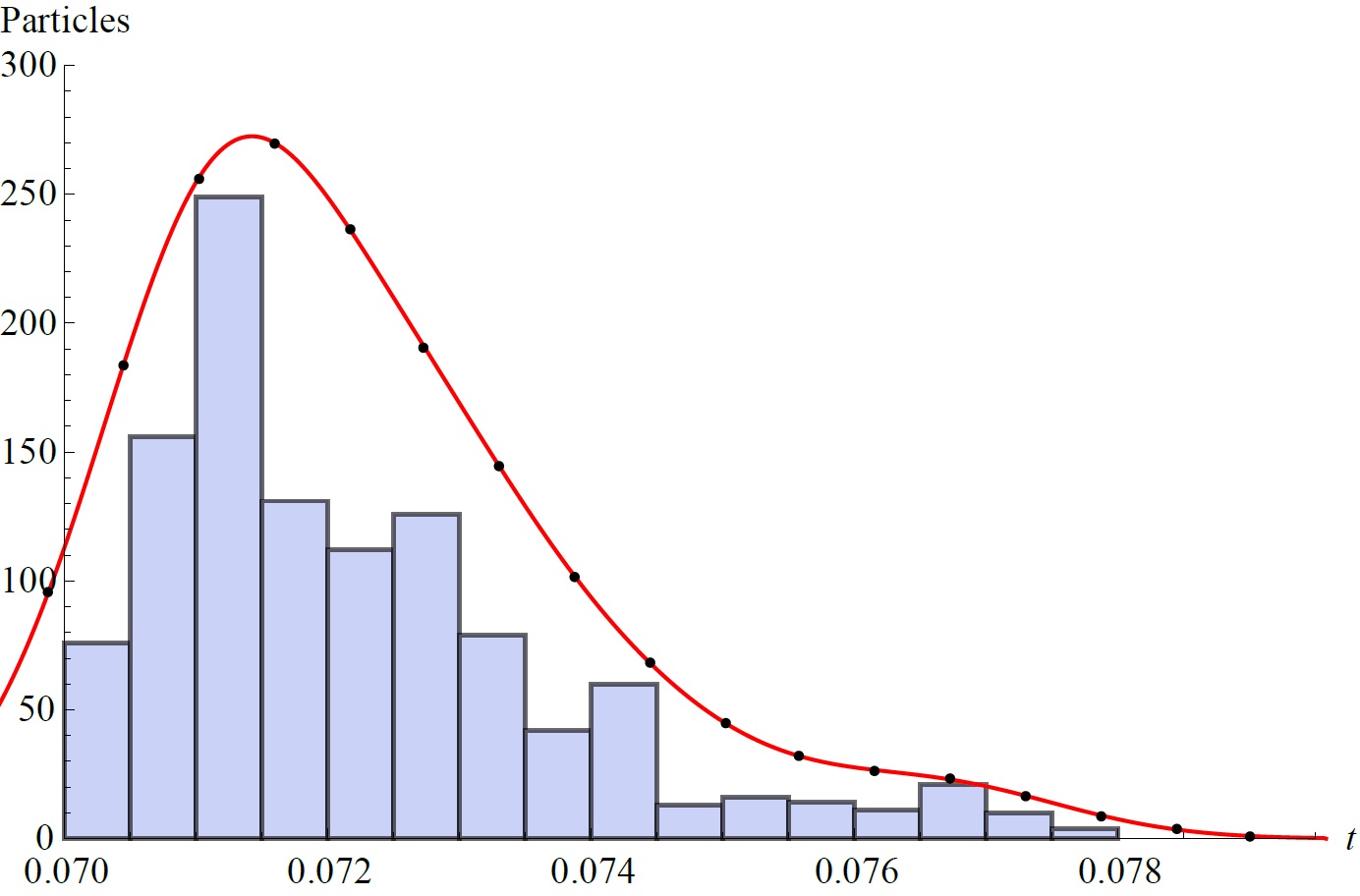}
    \caption{Time of arrival of particles to the screen located in $x=-350$.}
    \label{fig: Time of arrival}
\end{figure}

Another interesting feature of our model, due to its semiclassical nature, is the possibility to analyze the time of arrival or flight time, in quantum systems \cite{PhysRevA.77.014102}. For our double slit model we used similar initial conditions for classical and quantum variables to those used in actual experimental setups, but not so for initial velocities. We are able to obtain times of arrival for every configuration of our system, as shown in Fig.\ref{fig: Time of arrival}, although, in order to get more  experimentally interesting results, a more thorough analysis must be performed.

Our formulation can be employed in the study of complex quantum systems, such as the simulation of the double-slit experiment \cite{doi:10.1119/1.1858484, RevModPhys.60.1067}, the double-slit quantum eraser \cite{PhysRevA.65.033818}, the study of semiclassical quantum billiards \cite{PhysRevE.57.5397,DEALCANTARABONFIM2000129}, electron and neutron optical systems \cite{doi:https://doi.org/10.1002/3527602976.ch1,scully_zubairy_1997}, and several other applications.

\newpage

\printbibliography

\end{document}